%% file: main.tex
\documentclass{article}
\usepackage[T1]{fontenc} 
\usepackage[utf8]{inputenc} 
\usepackage{ismir,amsmath,cite,url}
\usepackage{graphicx}
\usepackage{color}
\usepackage{adjustbox}
\usepackage[table]{xcolor}
\usepackage{comment}

\usepackage{lineno}

\title{Between the AI and Me: Analysing Listeners' Perspectives on AI- and Human-Composed Progressive Metal Music}

\multauthor
{Pedro Sarmento \hspace{1cm} Jackson Loth \hspace{1cm} Mathieu Barthet \vspace{-0.4cm} } {\\Centre for Digital Music \\
 Queen Mary University of London\\
{\tt\small \{p.p.sarmento, j.j.loth, m.barthet\}@qmul.ac.uk}
}

\def\authorname{P. Sarmento, J. Loth, and M. Barthet}

\begin{document}

\maketitle

\input{Sections/abstract}
\input{Sections/intro}
\input{Sections/background}
\input{Sections/methodology}
\input{Sections/results}

\input{Sections/discussion}
\input{Sections/ethical}

\input{Sections/conclusion}
\input{Sections/acknowledgements}

\bibliography{Bibliography}

\end{document}

%% file: Sections/abstract.tex
\begin{abstract} 
Generative AI models have recently blossomed, significantly impacting artistic and musical traditions. Research investigating how humans interact with and deem these models is therefore crucial. Through a listening and reflection study, we explore participants' perspectives on AI- vs human-generated progressive metal, in symbolic format, using rock music as a control group. AI-generated examples were produced by ProgGP \cite{Loth2023ProgGP}, a Transformer-based model. We propose a mixed methods approach to assess the effects of generation type (human vs. AI), genre (progressive metal vs. rock), and curation process (random vs. cherry-picked). This combines quantitative feedback on genre congruence, preference, creativity, consistency, playability, humanness, and repeatability, and qualitative feedback to provide insights into listeners' experiences. A total of 32 progressive metal fans completed the study. Our findings validate the use of fine-tuning to achieve genre-specific specialization in AI music generation, as listeners could distinguish between AI-generated rock and progressive metal. Despite some AI-generated excerpts receiving similar ratings to human music, listeners exhibited a preference for human compositions. Thematic analysis identified key features for genre and AI vs. human distinctions. Finally, we consider the ethical implications of our work in promoting musical data diversity within MIR research by focusing on an under-explored genre.

\end{abstract}

%% file: Sections/intro.tex
\section{Introduction}

Recently, advancements in AI have resulted in generative models capable of creating remarkable musical pieces. This has been  particularly evident in the audio domain, with models such as Jukebox (OpenAI) \cite{Dhariwal2020}, MusicLM (Google) \cite{agostinelli2023musiclm}, AudioCraft/MusicGen (Meta) \cite{copet2024simple} and Stable Audio (Stability AI) \cite{evans2024fast}, to name a few. In contrast to audio generative models, which can produce complete, directly perceivable music with limited user input beyond specifying a prompt, symbolic music generation methods yield outputs that necessitate subsequent decoding and interpretation by performers and mixing by audio engineers before transforming them into music suitable for listening experiences. Unless automated using synthesis and machine mixing, by requiring human interpretation through performance and mixing, symbolic music rendering opens the door for the infusion of cultural and social elements. These elements become integral aspects of the final musical experience for listeners. 

One major controversy surrounding AI music generation models is their training on copyrighted data, often without consent nor royalty mechanisms. This raises concerns that AI-generated music could threaten artists' and musicians' income streams, amongst others \cite{barnett2023ethical}. These issues apply to both symbolic and audio AI music generation. However, symbolic approaches may pose less risk to artists' revenue streams since human musicians are still essential to the final product.

In this work, we focus on symbolic generative AI applied to progressive metal, which is considered a sub-genre of metal. Building on progressive rock's complex phrasing and odd time signatures, it incorporates a heavier focus on guitars and metal influences. The genre encompasses prominent bands such as Dream Theater, Between The Buried And Me\footnote{Used as inspiration for the title of this paper.}, and Meshuggah \cite{robinson2019exploration}. It is, however, relatively unexplored in academic literature, particularly in the context of AI music generation and MIR research \cite{wagner2010mean} \cite{hannan2019hearing}. 

\begin{figure}[h]
    \centering
    \includegraphics[width=1\columnwidth]{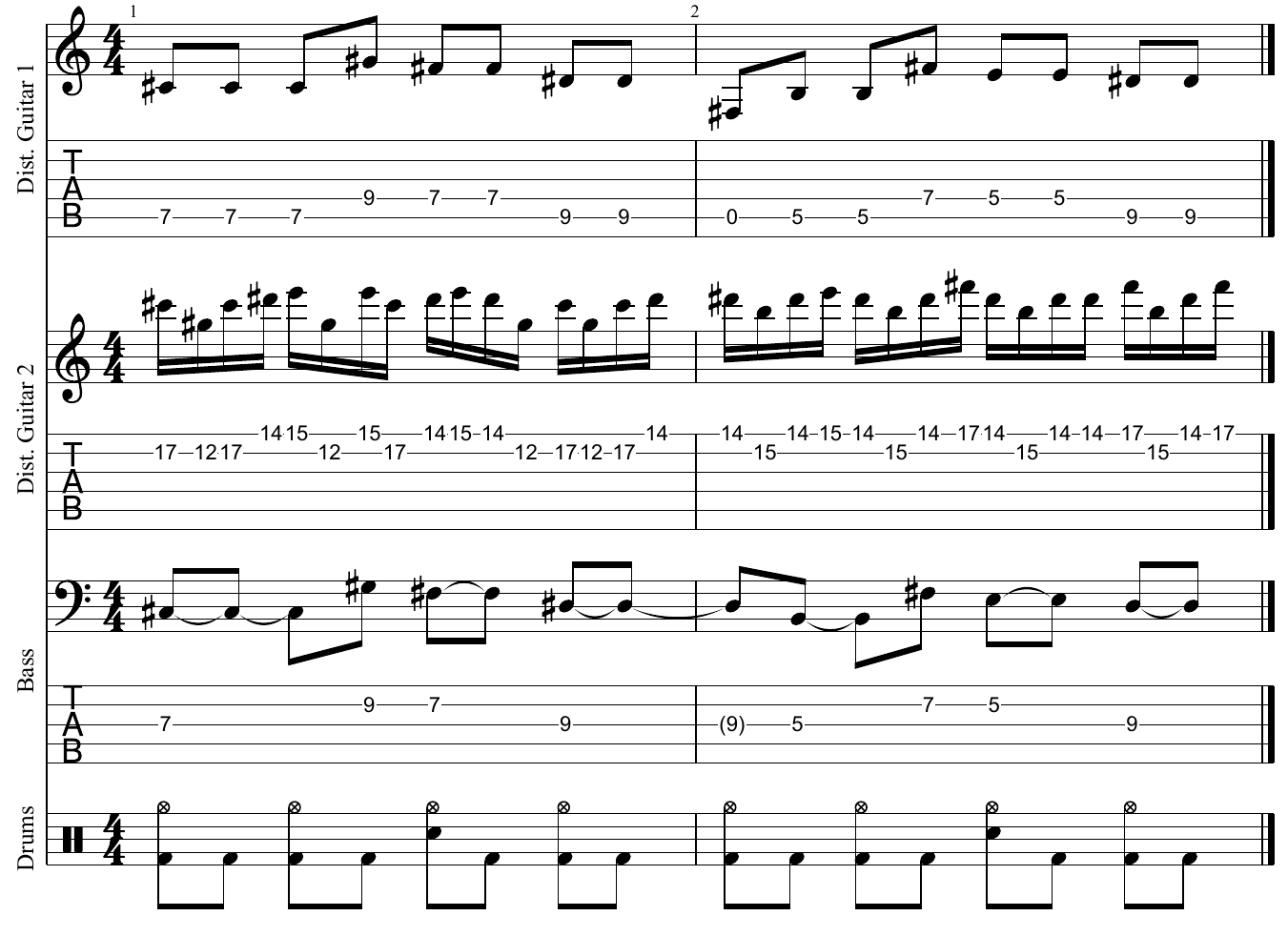}
    \caption{A screenshot from Guitar Pro of two measures from an AI-generated progressive metal song.}
    \label{fig:tabexample}
\end{figure}


Guitar tablature (see Figure \ref{fig:tabexample}) is a symbolic musical notation that translates guitar notes into fret and string numbers. Due to the genre's emphasis on guitar, progressive metal bands commonly use tablature to notate their compositions. Given that technical complexity is a large appeal of the genre, artists often sell their music in the form of tablatures for learning purposes through tablature publishing companies\footnote{As an example, Sheet Happens Publishing: \url{https://www.sheethappenspublishing.com/}}. Musicians from within the genre often use digital representations of tablatures and software like Guitar Pro\footnote{\url{https://www.guitar-pro.com/}} for dissemination of musical ideas and computer-assisted music making. 

We conducted a listening and reflection study to explore participants' perceptions of AI-generated progressive metal music. We used examples generated by ProgGP \cite{Loth2023ProgGP}, an AI model for multi-instrument guitar tablature creation (an example is shown in Figure \ref{fig:tabexample}). Overall, the contributions of this paper are: (1) a listening and reflective study methodology and questionnaire; (2) a subjective assessment of the capabilities of ProgGP for symbolic music generation, particularly in tablature format; (3) identification of compositional features of the progressive metal genre; (4) a critical analysis of AI-generated music through the lens of the progressive metal community; (5) an ethical reflection on musical data diversity in MIR, propelled by this study focusing on the underexplored progressive metal genre.

\vspace{-0.5cm}

%% file: Sections/background.tex
\vspace{-0.25cm}\section{Background}

\subsection{Symbolic Music Generation}

Music generation has seen an increase in popularity due to recent advances in deep learning \cite{sarmentoPerspectivesFutureSonic2021}, with many researchers utilizing techniques such as Recurrent Neural Networks (RNNs) \cite{Meade2019} \cite{Sturm2016}, Variational Autoencoders (VAEs) \cite{Tan2020}, Generative Adversarial Networks (GANs) \cite{Dong2018}, and Transformers \cite{AnnaHuang2019}. The Transformer model \cite{Vaswani2017},  known for its performance in natural language processing (NLP) tasks, has been adapted for generating symbolic piano music in Huang et al.'s Music Transformer \cite{AnnaHuang2019}, with Musenet \cite{christine_2019} and Pop Music Transformer \cite{Huang2020} further improving the approach. 

The field of guitar tablature generation gained significant momentum with the release of the DadaGP dataset \cite{Sarmento2021}. This dataset provides songs in two formats: GuitarPro, a popular tablature editing software, and a dedicated textual token format. This allows researchers to develop AI models that can both represent and generate music in tablature format. GTR-CTRL \cite{Sarmento2023} implements a Transformer-based model \cite{Huang2020}  for generating tablature that incorporates multiple instruments. It offers control over instrumentation (inst-CTRL) and musical genre (genre-CTRL). ProgGP \cite{Loth2023ProgGP}, the model used in this study, focuses specifically on the progressive metal genre (see Figure \ref{fig:tabexample} and description in Section \ref{proggp-qualitative}). LooperGP \cite{Adkins2023} adapts the method to generate loopable music excerpts, making it applicable e.g. for live coding performances. By fine-tuning the model on the music of four iconic guitar players, ShredGP \cite{Sarmento2023ShredGP} demonstrates its ability to replicate specific styles.

\vspace{-0.2cm}\subsection{Subjective Evaluation of AI-Generated Music}
\label{sec:subjective_eval}

Objective computational measures can provide an initial assessment of AI-generated music quality \cite{Yang:EvaluationofMusicGen}. However, often they struggle to capture the subtleties needed to judge their aesthetic merit.  The combination of objective computational measures with subjective human evaluations provides a more holistic understanding of AI-generated music. Listening tests typically involve ranking or scoring AI- and human-generated stimuli according to several metrics to gain a more comprehensive understanding of perceived quality. This often involves comparing outputs from different models with the established reference (the known ideal or benchmark). Metrics used to assess AI music vary from general attributes such as \textit{musicality} \cite{AnnaHuang2019}, \textit{liking} \cite{deguernel2022investigating}\cite{renPopMAGPopMusic2020}\cite{Huang2020}\cite{daiControllableDeepMelody2021a}, \textit{pleasantness} \cite{Dong2018a}, \textit{richness} \cite{luMeloFormGeneratingMelody2022}, to more specific qualities such as \textit{consistency} \cite{wuJukeDrummerConditionalBeataware2022}, or \textit{structural/stability} properties \cite{wuJukeDrummerConditionalBeataware2022}. Whereas ranking involves sorting the different stimuli along a given dimension, scoring tasks commonly rely on 5- or 7-point Likert items \cite{jiSurveyDeepLearning2023}\cite{yang2017midinet}. A musical Turing test, similar to the original Turing test, is designed to assess a machine's ability to exhibit human-level musical creation features. In such tests, participants attempt to distinguish between human- and AI-generated music \cite{hadjeres2017deepbach}. To assess AI-generated music, we employ a mixed methods approach, combining quantitative and qualitative data from a listening and reflection study. This approach, common in music perception and HCI research (e.g. \cite{Yang2023Emotion}), allows for a deeper understanding of the problem. While listening tests enable us to better understand human perception of AI-generated music, they are not without limitations. These limitations include listener fatigue, potential biases due to stimuli or participant selection. Additionally, they may lack sufficient statistical power to generalize the findings to a broader population.

\vspace{-0.3cm}

%% file: Sections/methodology.tex
\section{Methodology}
\label{proggp-qualitative}
We used a mixed methods listening and reflective study to assess AI music, with an ethical approval from the Queen Mary Ethics of Research Committee. All data was collected anonymously. The study took around 1h to complete and participants were compensated with a £10 Amazon voucher.


We evaluated AI-generated progressive metal music from the ProgGP model \cite{Loth2023ProgGP}, a Transformer-XL model trained on the DadaGP dataset \cite{Sarmento2021} and fine-tuned on a progressive metal corpus, by comparing it to human-composed progressive metal pieces. We compare two ways of choosing AI music: picking songs at random and subjectively choosing the ``best'' ones (cherry-picking) through active listening. Additionally, we compare the ProgGP model's outputs with rock music generated by the genre-CTRL model \cite{Sarmento2023}, a similar model conditioned on the rock genre. Human-composed rock music serves as another control group in this comparison.

\subsection{Hypotheses}

Our study tests the following hypotheses: 
 \begin{itemize}\vspace{-0.175cm}
    \item $\boldsymbol{H_1}$\textbf{: Human-composed music obtains better scores than AI-generated music.} We compare AI- and human-generated music along the following dimensions: preference, creativity, consistency, playability and repeatability. 
    \vspace{-0.2cm}
    \item $\boldsymbol{H_2}$\textbf{: AI-generated and human-composed music can be distinguished.} This hypothesis is linked to the musical Turing test.\vspace{-0.2cm}
    \item $\boldsymbol{H_3}$\textbf{: AI-generated music matches the genre used for model conditioning.} The ability of the model to specialize in a specific genre (progressive metal).\vspace{-0.2cm}
    \item  $\boldsymbol{H_4}$\textbf{: Cherry-picked AI-generated music is preferred to randomly chosen AI-generated music.} We hypothesize that picking examples by hand leads to better performance than random selection.
\end{itemize}  

\vspace{-0.2cm}

\vspace{-0.3cm}

\subsection{Stimuli}


The stimuli were rendered using Guitar Pro 7, a software for playing/editing  digital guitar tablatures. The human-composed music was obtained using publicly available transcriptions of  progressive metal and rock songs hosted on Songsterr\footnote{\url{https://www.songsterr.com/}}, a website hosting Guitar Pro tablatures, as well as from the DadaGP dataset \cite{Sarmento2021}. All the examples were trimmed to 15 seconds, and rendered as WAV files using the default virtual instruments in Guitar Pro 7. They were further loudness-normalized \cite{steinmetz2021pyloudnorm}. The study comprised 60 stimuli\footnote{\sloppy\url{https://drive.google.com/drive/folders/1-PVPXNCMu73ICfNf0qlwxzdVpNxrWIVL?usp=sharing}} broken down into the following six groups with 10 examples per group: \textbf{progcp} (\textbf{prog}ressive metal examples generated using ProgGP \textbf{c}herry-\textbf{p}icked), \textbf{progrand} (\textbf{prog}ressive metal examples generated with ProgGP, \textbf{rand}omly selected), \textbf{proghum} (\textbf{prog}ressive metal examples from the dataset used to fine-tune ProgGP, \textbf{hum}an-generated, randomly selected), \textbf{rockcp} (\textbf{rock} examples generated using genre-CTRL prompted for rock, \textbf{c}herry-\textbf{p}icked), \textbf{rockrand} (\textbf{rock} examples generated using genre-CTRL prompted for rock, \textbf{rand}omly selected), and \textbf{rockhum} (from \textbf{rock} examples in the dataset used for genre-CTRL, \textbf{hum}an-generated, randomly selected). The AI-generated stimuli were selected out of a corpus of 200 compositions from each genre.

\subsection{Participants}

We recruited participants familiar with progressive metal as we wanted to involve domain experts capable of identifying differences between rock and progressive metal. To this end, we advertised the call for participants on the \textit{r/progmetal} sub-forum from the Reddit platform. This community comprises progressive metal aficionados\footnote{Available at: \sloppy\url{https://www.reddit.com/r/progmetal/}}. 26 participants were gathered from this forum. We recruited six additional progressive metal fans within our department, for a total of 32 participants. Their age distribution was $29\pm5$ years old, with 27 males and 5 females. The participants had an average Gold-MSI score of $81.41$, indicating average level of musical sophistication compared to results of previous studies \cite{goldsmith2014}. \vspace{-0.3cm}

\subsection{Procedure}

Participants first went through a familiarization stage containing two excerpts, followed by the main task during which musical excerpts were presented in random order to minimise potential order effects. Participants were instructed to focus on the quality of the composition and not on the quality of the virtual instruments or the music production mix. For each excerpt, participants had to listen to the stimulus, report their familiarity, and answer the following questions using 7-point Likert items: $\boldsymbol{Q_1}$ (``This composition features the qualities of the progressive metal genre.''), $\boldsymbol{Q_2}$ (``This composition features the qualities of the rock genre.''), $\boldsymbol{Q_3}$ (``I like this composition.''), $\boldsymbol{Q_4}$ (``This composition is creative.''), $\boldsymbol{Q_5}$ (``This composition is consistent.''), $\boldsymbol{Q_6}$ (``This composition is playable.''), $\boldsymbol{Q_7}$ (``This composition was generated using AI.'') and $\boldsymbol{Q_8}$ (``This composition is repetitive.''). Once the participants finished rating all excerpts, they were presented with a post-task questionnaire to assess their reasoning when distinguishing between genres as well as between AI- and human-composed excerpts (see Section \ref{sec:thematicanalysis}). \vspace{-0.3cm}

%% file: Sections/results.tex
\section{Results}\vspace{-0.2cm}

\subsection{Listening Test}
\label{sec:listeningtest}

\input{Sections/results_LT}

\vspace{-0.2cm}\subsection{Thematic Analysis}
\label{sec:thematicanalysis}

\input{Sections/results_TA}

%% file: Sections/results_LT.tex
\begin{figure*}[h]
    \centering
    \includegraphics[width=1\textwidth]{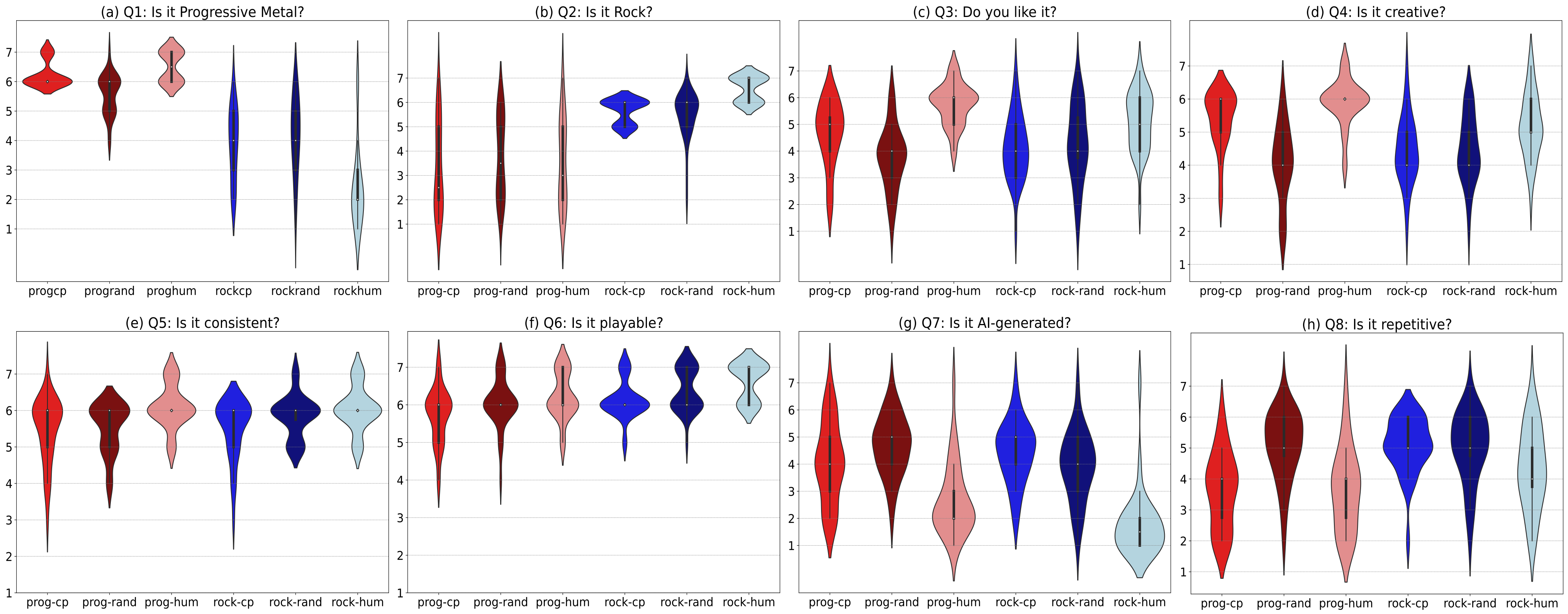}
    \caption{Violin plots of answers to Likert items for $Q_1$ to $Q_8$, in plots (a) to (h), respectively, providing an estimation of the probability density function of the data. Horizontal axis represents the different groups of stimuli. Vertical axis reports the 7-point Likert ratings from 1 (Strongly Disagree) to 7 (Strongly Agree).}
    \label{fig:violinplots}
\end{figure*}

We visualize Likert item answers using violin plots in Figure \ref{fig:violinplots}. We conducted statistical analyses investigating the effects of the music creation process (six levels: \textbf{progcp}, \textbf{progrand}, \textbf{proghum}, \textbf{rockcp}, \textbf{rockrand}, \textbf{rockhum}) on several dependent variables (preference, creativity, consistency, playability, repeatability, humanness, genre congruency, and AI curation method, where relevant). Because we employed a within-participant design with repeated measures, and the collected data are ordinal, we used the non-parametric Friedman test. We use a Type I error $\alpha$ of 0.05; results are presented in Table \ref{tab:friedman}. The Friedman test was followed by post-hoc pairwise Wilcoxon tests, using a Bonferroni-adjusted $\alpha$ level of $.0033$ ($.05/15$). This enables us to compare two generation types (AI vs. human), two genres (progressive metal vs. rock), and two AI selection methods (random vs. cherry-picked). Results are presented in Table \ref{tab:wilcoxon}. For question about AI-generated music (Q7), we excluded responses (345 out of 1,920) where participants indicated prior song familiarity. \vspace{-0.3cm}

\input{Tables/friedman}
\input{Tables/wilcoxon}

%% file: Tables/friedman.tex
\begin{table}
\begin{center}
\begin{adjustbox}
{width=\columnwidth}   
\begin{tabular}{c|c|c|c}
    \hline
    \hline
          \textbf{Question} & \textbf{Friedman Test Statistic} & $\boldsymbol{p}$\textbf{-value} & \textbf{Significance} \\
\hline
\hline
        $Q_1$ & $\chi^2(5)=136.90$ & $8.14 \times 10^{-28}$ & ***\\
        $Q_2$ & $\chi^2(5)=110.09$ & $3.90 \times 10^{-22}$  & ***\\
        $Q_3$ & $\chi^2(5)=77.56$ & $2.72 \times 10^{-15}$ & ***\\
        $Q_4$ & $\chi^2(5)=88.54$ & $1.36 \times 10^{-17}$ & ***\\
        $Q_5$ & $\chi^2(5)=42.50$ & $4.67 \times 10^{-8}$ & ***\\
        $Q_6$ & $\chi^2(5)=55.47$ & $1.04 \times 10^{-10}$ & ***\\
        $Q_7$ & $\chi^2(5)=51.59$ & $6.53 \times 10^{-10}$ & ***\\
        $Q_8$ & $\chi^2(5)=79.20$ & $1.23 \times 10^{-15}$ & ***\\ 
\hline
\hline
    \end{tabular}
\end{adjustbox}
\end{center}
\caption{Friedman test results investigating the effect of the creation method for each question ($Q_1$ to $Q_8$).}
\label{tab:friedman}
\end{table}

%% file: Tables/wilcoxon.tex
\begin{table*}
\begin{center}
\begin{adjustbox}
{width=\textwidth}\begin{tabular}{c|c|c|c|c|c|c|c|c|c}
\hline
\hline
\textbf{Group 1} & \textbf{Group 2} & $\boldsymbol{Q_1}$ & $\boldsymbol{Q_2}$& $\boldsymbol{Q_3}$& $\boldsymbol{Q_4}$& $\boldsymbol{Q_5}$& $\boldsymbol{Q_6}$& $\boldsymbol{Q_7}$& $\boldsymbol{Q_8}$\\
\hline
\hline
rockrand & rockhum & \cellcolor{green!35}$1.10 \times 10^{-4}$ & \cellcolor{green!35}$2.63 \times 10^{-5}$ & $0.26$ & $0.15$ & $4.29$ & $1.70$ & $0.44$ & $0.55$ \\
\hline
rockcp & rockrand & $6.42$ & $1.15 \times 10^{1}$ & $8.25$ & $1.32 \times 10^{1}$ & $3.00$ & $1.70$ & $1.30$ & $1.36 \times 10^{1}$ \\
\hline
rockcp & rockhum & \cellcolor{green!35}$3.96 \times 10^{-4}$ & \cellcolor{green!35}$1.43 \times 10^{-5}$ & $1.78 \times 10^{-2}$ & $7.52 \times 10^{-2}$ & $0.35$ & $2.30 \times 10^{-2}$ & $7.22 \times 10^{-3}$ & $0.19$ \\
\hline
progrand & proghum & \cellcolor{green!35}$5.79 \times 10^{-5}$ & $7.09$ & \cellcolor{green!35}$2.19 \times 10^{-7}$ & \cellcolor{green!35}$2.46 \times 10^{-7}$ & $9.24 \times 10^{-2}$ & $5.16$ & \cellcolor{green!35}$2.94 \times 10^{-6}$ & \cellcolor{green!35}$8.43 \times 10^{-4}$ \\
\hline
progrand & rockcp & \cellcolor{green!35}$3.78 \times 10^{-6}$ & \cellcolor{green!35}$2.17 \times 10^{-4}$ & $4.24$ & $4.42$ & $1.39 \times 10^{1}$ & $1.08 \times 10^{1}$ & $9.08$ & $1.40 \times 10^{1}$ \\
\hline
progrand & rockrand & \cellcolor{green!35}$1.86 \times 10^{-4}$ & \cellcolor{green!35}$1.36 \times 10^{-3}$ & $1.79$ & $3.60$ & $2.32$ & $0.97$ & $0.48$ & $1.26 \times 10^{1}$ \\
\hline
progrand & rockhum & \cellcolor{green!35}$3.45 \times 10^{-8}$ & \cellcolor{green!35}$7.61 \times 10^{-9}$ & \cellcolor{green!35}$5.47 \times 10^{-4}$ & $4.80 \times 10^{-3}$ & $0.24$ & $1.30 \times 10^{-2}$ & \cellcolor{green!35}$1.40 \times 10^{-3}$ & $0.34$ \\
\hline
progcp & progrand & $7.78 \times 10^{-3}$ & $3.56$ & $3.69 \times 10^{-2}$ & \cellcolor{green!35}$1.69 \times 10^{-4}$ & $1.44 \times 10^{1}$ & $1.65$ & $2.10$ & \cellcolor{green!35}$3.62 \times 10^{-4}$ \\
\hline
progcp & proghum & $0.80$ & $9.36$ & \cellcolor{green!35}$3.16 \times 10^{-3}$ & $0.25$ & $0.14$ & $0.22$ & $5.60 \times 10^{-3}$ & $1.44 \times 10^{1}$ \\
\hline
progcp & rockcp & \cellcolor{green!35}$9.29 \times 10^{-10}$ & \cellcolor{green!35}$3.79 \times 10^{-5}$ & $0.67$ & \cellcolor{green!35}$3.16 \times 10^{-3}$ & $1.36 \times 10^{1}$ & $0.64$ & $4.71$ & \cellcolor{green!35}$7.25 \times 10^{-5}$ \\
\hline
progcp & rockrand & \cellcolor{green!35}$2.38 \times 10^{-8}$ & \cellcolor{green!35}$1.98 \times 10^{-4}$ & $3.56$ & $8.38 \times 10^{-3}$ & $2.56$ & $1.78 \times 10^{-2}$ & $1.07 \times 10^{1}$ & \cellcolor{green!35}$7.31 \times 10^{-4}$ \\
\hline
progcp & rockhum & \cellcolor{green!35}$2.81 \times 10^{-9}$ & \cellcolor{green!35}$1.63 \times 10^{-8}$ & $2.59$ & $3.64$ & $0.33$ & \cellcolor{green!35}$1.24 \times 10^{-4}$ & $0.34$ & $0.60$ \\
\hline
proghum & rockcp & \cellcolor{green!35}$4.11 \times 10^{-10}$ & \cellcolor{green!35}$2.72 \times 10^{-5}$ & \cellcolor{green!35}$8.83 \times 10^{-6}$ & \cellcolor{green!35}$3.92 \times 10^{-6}$ & $0.14$ & $7.92$ & \cellcolor{green!35}$3.00 \times 10^{-5}$ & \cellcolor{green!35}$1.98 \times 10^{-4}$ \\
\hline
proghum & rockrand & \cellcolor{green!35}$3.50 \times 10^{-9}$ & \cellcolor{green!35}$2.17 \times 10^{-4}$ & \cellcolor{green!35}$5.31 \times 10^{-4}$ & \cellcolor{green!35}$9.47 \times 10^{-6}$ & $2.32$ & $5.69$ & $5.06 \times 10^{-3}$ & \cellcolor{green!35}$1.61 \times 10^{-3}$ \\
\hline
proghum & rockhum & \cellcolor{green!35}$8.49 \times 10^{-10}$ & \cellcolor{green!35}$1.93 \times 10^{-8}$ & $0.37$ & $2.40 \times 10^{-2}$ & $1.08 \times 10^{1}$ & $0.25$ & $3.00$ & $0.65$ \\

\hline
\hline
\end{tabular}
\end{adjustbox}
\end{center}
\caption{Pairwise post-hoc wilcoxon test  results for each question. Each cell indicates $p$-value, while green cells indicates statistical significance (i.e. with Bonferroni correction $p<0.0033$).}
\label{tab:wilcoxon}
\end{table*}

%% file: Sections/results_TA.tex
We performed a thematic analysis \cite{braun2006using} of answers to post-task questions to better understand the thought process of participants' decisions during the study. Multiple themes were obtained from the responses, and results are  ordered by number of codes within each theme (in parentheses next to each theme, indicating number of occurences).

\vspace{-0.3cm}\subsubsection{What features made you identify excerpts as progressive metal?}
\label{sec:TA_prog}

\noindent \textbf{Complexity (40):} A huge emphasis was put on the complexity of a composition, particularly the rhythmic but also the harmonic and melodic complexity. Uncommon and changing time signatures were mentioned by roughly half of the participants. The difficulty of playing a composition was also a very common answer. 

\noindent\textbf{Composition/style (38):} Many compositional and stylistic elements were seen as particularly relevant to the genre, such as aggressiveness, speed and atmosphere. A sense of cohesion is important, with ``clear and distinct ideas glued together''. The composition should be experimental, with creative rhythms, unique segments and interesting harmonic choices. Dissonant melodies, arpeggios, metal drum patterns and guitar specific techniques such as ``chugs'' are also deemed as important.

\noindent\textbf{Instrumentation (7):} Participants mention unique instruments and extended range guitars being particularly indicative of the genre.

\subsubsection{What features made you identify excerpts as Rock?} \label{sec:TA_rock} 

\noindent\textbf{Musical structure/composition (24):} These excerpts were repetitive and had slower tempos, utilizing a question and answer structure and accents on beats two and four. They were also generally soft and not particularly aggressive.

\noindent\textbf{Simple/straightforward (23):} The excerpts identified as rock were seen as simplistic, using simple drums, melodies, chord progressions and particularly 4/4 time signatures. These songs had straightforward grooves and generic solos.  

\noindent\textbf{Guitar techniques (14):} Many techniques were seen as specific to the rock genre such as the use of the pentatonic scale, open chords, power chords and double stops. Participants noted a clear blues inspiration in the guitar playing.

\noindent\textbf{Instrumentation (11):} The rock genre was seen as guitar driven, with guitars and bass parts being separated. The drums were generally synchronized with the guitar and emphasized the hi-hat cymbals.

\subsubsection{What made you identify excerpts as being composed using AI? } 
\label{sec:TA_AI}


\noindent\textbf{Something ``off'' about the composition (40):} A major theme involved participants having some feeling of unease about the composition. Preference for human-composed music might be attributed to a perceived lack of qualities often associated with human creation, such as ``soul" and creativity, or the inability to emulate human-like musical performance (playability). Many participants noted that some compositions lacked a sense of cohesiveness and consistency, or even sounded random, with odd note choices and bass lines which did not make sense. Participants also felt there was too much complexity, but also not enough of particular types of complexity (e.g. harmonic complexity).

\noindent\textbf{Repetition (14):} This theme specifically refers to negatively perceived repetition. Many described excerpts being overly repetitive or repeating ``musically uninteresting or unsatisfying phrases''.

\noindent\textbf{Uninteresting/simple (8):} Participants describe boring and generic riffs as well as simplistic and bland drum patterns.

\noindent\textbf{Melody (7):} A lack of interest or satisfaction with melodies was mentioned, specifically melodies that ``run too long and miss their resolution'' and ``do not seem go anywhere''.

\subsubsection{What made you identify excerpts as being composed by humans?}
\label{sec:TA_human}

\noindent\textbf{Well-composed (36):} A sense of cohesion and consistency throughout the instrumentation and musical ideas was a popular reason for identifying an excerpt as human. Many also mentioned musical choices which feel deliberate and intentional. In general, compositions which felt natural, predictable, and emotionally satisfying were seen as more human.

\noindent\textbf{Human-qualities (10):} Certain qualities were perceived as more human, such as creativity, ``soul'', and playability. The use of music theory, as well as breaking the rules of music theory were also mentioned.

%% file: Sections/discussion.tex
\section{Discussion}


\vspace{-0.1cm}\subsection{$\boldsymbol{H_1}$: Human-composed music obtains better scores than AI-generated music}

Human-composed progressive metal (\textbf{proghum}) was significantly preferred to all the other AI-generated groups (see $Q_3$ in Table 2). However, this could be due to participants all being progressive metal fans. Our findings suggest that a Turing test style approach may have limitations in evaluating generative models. While participants struggled to distinguish AI-generated from human-composed music, they still preferred the human compositions.



Participants' evaluations used more negative language (e.g., 'repetitive') to describe AI compositions and more positive language for human compositions (see Sections \ref{sec:TA_AI} and \ref{sec:TA_human}). One might naturally expect significant differences in responses to the listening experiment between the AI-generated and human-composed music stimuli groups. While this is true for the randomly selected AI-generated progressive metal group (\textbf{progrand}), it does not hold for the rock groups as well as the cherry-picked AI-generated progressive metal (\textbf{progcp}) group. However, the violin plots (see Figure \ref{fig:violinplots}) do show the human-composed groups to generally have a better mean and smaller variance. $Q_5$ (``This composition is consistent'') and $Q_6$ (``This composition is playable'') saw no or few significant differences between stimuli groups. One of the negatively framed indicators of AI compositions was repetition. The cherry-picked and human-composed progressive metal groups were both significantly different to every group other than each other and the human-composed rock (\textbf{rockhum}) group in the responses to $Q_8$ (``This composition is repetitive''). Figure \ref{fig:violinplots} also shows the responses in these groups trending toward not repetitive, while the others trend closer to repetitive. The \textbf{rockhum} group was not significantly different to any of the other groups, both AI and human-composed. While the level of repetition in AI-generated excerpts may be roughly similar to human-composed excerpts in their respective genre (with the exception of \textbf{progrand}), it is possible that repetition quality is different between AI and human-composed excerpts. Overall, we can conclude that the test shows strong evidence for $H_1$ in terms of preference, but not necessarily for the other dimensions. 

\vspace{-0.4cm}\subsection{$\boldsymbol{H_2}$: AI-generated and human-composed music can be distinguished}

Figure \ref{fig:violinplots} shows a large variance in the responses for $Q_7$ (``This composition was generated using AI'') for the AI generated stimuli groups, as well as numerous classification errors. The human stimuli groups also show classification errors, though less when compared to the AI stimuli groups. The \textbf{proghum} stimuli group was significantly different from both the cherry-picked rock (\textbf{rockcp}) and \textbf{progrand} groups. The \textbf{rockhum} group was only significantly different to the \textbf{progrand} group. The \textbf{progcp} and randomly selected AI-generated rock (\textbf{rockrand}) groups were not found to be significantly different to either of the human-composed groups. Ultimately, this is evidence against $H_2$, though there seems to be some dependence on the model used and the samples selected from that model.

\vspace{-0.3cm}\subsection{$\boldsymbol{H_3}$: AI-generated music matches the genre used for model conditioning}

The responses to $Q_1$ and $Q_2$ (see Figure \ref{fig:violinplots} (a) and (b)) show a clear ability of participants to distinguish between the genres of progressive metal and rock, supporting $H_3$. This is expected given that participants described the genres very differently to each other (see Sections \ref{sec:TA_prog} and \ref{sec:TA_rock}). Of the progressive metal stimuli groups, only \textbf{progrand} and \textbf{proghum}, differed significantly, suggesting that at least the human curated AI-generated progressive metal stimuli (\textbf{progcp}) have features of similar quality to those of the human-generated group. The same cannot be said of the rock samples, though Figure \ref{fig:violinplots} (b) shows the mean ratings in the rock groups are clearly higher than neutral (rated as 4), indicating that they identified the samples as rock. This may suggest that ProgGP excels in comparison to genre-CTRL in generating musical examples in its target genre. However, genre-CTRL, being trained on a wider range of styles (rock, punk, metal, classical, folk), could theoretically generate music in various styles, unlike ProgGP which is limited to its training genre.

\vspace{-0.3cm}\subsection{$\boldsymbol{H_4}$: Cherry-picked AI-generated music is preferred over randomly chosen AI-generated music}

At the surface level, the cherry-picked and randomly selected stimuli groups do not seem to have many differences. The groups in the rock genre have no significant differences between them in any of the questions, and the progressive metal groups only yield significant differences for $Q_4$ (``This composition is creative''). However, we observe that there are several questions with significant differences between the \textbf{progrand} and \textbf{proghum} groups, while the \textbf{progcp} and \textbf{proghum} groups only differ for $Q_3$. This seems to indicate that the \textbf{progcp} stimuli have more in common with the human-composed excerpts than the randomly selected ones do in the tested features. Additionally, the difference seen in $Q_4$ between the AI-generated progressive metal groups concerns creativity, shown to be an indicator of human composition in Section \ref{sec:TA_human}. It is difficult to make any definite conclusions about $H_4$ given these results, but there seems to be some weak evidence for it in the progressive metal genre.

\vspace{-0.25cm}\subsection{Study Limitations}

The study is bounded by the number of participants ($32$) and an unbalanced gender distribution. Moreover, the stimuli were only $15$ seconds long each, meaning that participants could not judge any long-term compositional features. Finally, the responses discussed in the study focus on compositional features and discard expressive and timbre-related aspects.

%% file: Sections/ethical.tex
\section{Ethics of Musical Data Diversity}


The broader topic of diversity within MIR is debated by Born in \cite{Born2020}, in which the author highlights points such as (1) the demographics within the field, (2) the nature of the music that is commonly researched, questions (3) the applicability of scientific work to a broader, more diverse, corpus of music, and (4) how to better stir MIR research towards a more encompassing music economy. Of particular relevance to this reflection is (2), closely linked to the concerns on musical data diversity. Despite efforts towards research concerning traditional, folk or ethnic music, MIR is still predominately based on the mainstream popular music that follows a ``western'' tradition \cite{Gomez2013ComputationalEP}. Moreover, by referring to an ISMIR keynote by Seeger \cite{Seeger2003IFoundIt}, Sturm et al. \cite{sturmCopyright2019} point out that even within ``western'' music, there seems to be an emphasis on US pop music and European classical music. For automatic symbolic music generation, a closer look at the most commonly used datasets in the MIR community \cite{Dong2020} suggests that styles such as western classical, pop and jazz music, often modelled using piano when dealing with single instrument systems, are a recurrent practice within the field. Following from these premises, it is important to first clarify that ProgGP is still grounded in the western music tradition, and to acknowledge this as a limitation given the musical data diversity concerns explained before. However, its musical style can be said to emphasize content that diverges from the \textit{mainstream popular music} landscape. This begged the question: can the stylistic biases in the outputs from ProgGP contribute to a wider context of data diversity within MIR research? 
We argue that releasing training data for specific genres, exemplified by the release of data for fine-tuning ProgGP \cite{Loth2023ProgGP}, is a step towards a more musically diverse MIR. This, along with publishing studies to understand underexplored genres and their challenges, and fostering interaction with stakeholders like the progressive metal community (as in this paper), can significantly contribute to this goal. 
We propose that guitar tablature can enhance musical diversity in MIR. Unlike MIDI, the dominant format, tablature excels at representing string instrument-specific expressive techniques, expanding the scope of representable music within the field. \vspace{-0.2cm}




%% file: Sections/conclusion.tex
\vspace{-0.3cm}\section{Conclusion}

We conducted a listening and reflective study which examined listeners perspectives on the quality of symbolic AI-generated compositions in the rock and progressive metal genres. The study provided both a subjective evaluation of recent Transformer-based music generation models and an exploration of listeners' perceptions of AI and human compositions. We found that participants preferred human-composed music over AI-generated music, though they were generally not able to fully distinguish between AI- and human-composed music. Participants were able to distinguish between the two genres well. Cherry-picked examples in the progressive metal genre were rated similarly to the human-composed examples in several compositional metrics despite not being liked as much. With this methodology, we hope our work helps researchers better evaluate their generative models using a mixed methods approach through a listening and reflective study, as well as show the merit in increasing musical data diversity within MIR.

%% file: Sections/acknowledgements.tex
\section{Acknowledgements}
This work is supported by the EPSRC UKRI Centre for Doctoral Training in Artificial Intelligence and Music (Grant no. EP/S022694/1).

%% file: main.bbl
\begin{thebibliography}{10}
\providecommand{\url}[1]{#1}
\csname url@samestyle\endcsname
\providecommand{\newblock}{\relax}
\providecommand{\bibinfo}[2]{#2}
\providecommand{\BIBentrySTDinterwordspacing}{\spaceskip=0pt\relax}
\providecommand{\BIBentryALTinterwordstretchfactor}{4}
\providecommand{\BIBentryALTinterwordspacing}{\spaceskip=\fontdimen2\font plus
\BIBentryALTinterwordstretchfactor\fontdimen3\font minus \fontdimen4\font\relax}
\providecommand{\BIBforeignlanguage}[2]{{%
\expandafter\ifx\csname l@#1\endcsname\relax
\typeout{** WARNING: IEEEtran.bst: No hyphenation pattern has been}%
\typeout{** loaded for the language `#1'. Using the pattern for}%
\typeout{** the default language instead.}%
\else
\language=\csname l@#1\endcsname
\fi
#2}}
\providecommand{\BIBdecl}{\relax}
\BIBdecl

\bibitem{Loth2023ProgGP}
J.~Loth, P.~Sarmento, C.~Carr, Z.~Zukowski, and M.~Barthet, ``{ProgGP: From GuitarPro Tablature Neural Generation To Progressive Metal Production},'' in \emph{The 16th International Symposium on Computer Music Multidisciplinary Research}, 2023.

\bibitem{Dhariwal2020}
\BIBentryALTinterwordspacing
P.~Dhariwal, H.~Jun, C.~Payne, J.~W. Kim, A.~Radford, and I.~Sutskever, ``{Jukebox: A Generative Model for Music},'' 2020. [Online]. Available: \url{https://github.com/openai/jukebox.}
\BIBentrySTDinterwordspacing

\bibitem{agostinelli2023musiclm}
A.~Agostinelli, T.~I. Denk, Z.~Borsos, J.~Engel, M.~Verzetti, A.~Caillon, Q.~Huang, A.~Jansen, A.~Roberts, M.~Tagliasacchi \emph{et~al.}, ``Music{LM}: Generating {M}usic from {T}ext,'' \emph{arXiv preprint arXiv:2301.11325}, 2023.

\bibitem{copet2024simple}
J.~Copet, F.~Kreuk, I.~Gat, T.~Remez, D.~Kant, G.~Synnaeve, Y.~Adi, and A.~Défossez, ``{Simple and Controllable Music Generation},'' in \emph{37th Conference on Neural Information Processing Systems (NeurIPS)}, 2023.

\bibitem{evans2024fast}
Z.~Evans, C.~Carr, J.~Taylor, S.~H. Hawley, and J.~Pons, ``{Fast Timing-Conditioned Latent Audio Diffusion},'' in \emph{Proceedings of the 41st International Conference on Machine Learning}, 2024.

\bibitem{barnett2023ethical}
J.~Barnett, ``The ethical implications of generative audio models: A systematic literature review,'' in \emph{Proceedings of the AAAI/ACM Conference on AI, Ethics, and Society (AIES '23)}, 2023, p.~10, 1 figure.

\bibitem{robinson2019exploration}
D.~Robinson, ``{An Exploration of the Various Compositional Approaches to Modern Progressive Metal},'' MA Thesis, University of Huddersfield, 2019.

\bibitem{wagner2010mean}
J.~Wagner, \emph{{Mean Deviation: Four Decades of Progressive Heavy Metal}}.\hskip 1em plus 0.5em minus 0.4em\relax Bazillion Points Books, 2010.

\bibitem{hannan2019hearing}
C.~Hannan, ``{Hearing Form in Progressive Metal: Motivic Return, Genre Borrowing, and Sonata Form in Between the Buried and Me’s Parallax II’},'' MA Thesis, Columbia University New York, 2019.

\bibitem{sarmentoPerspectivesFutureSonic2021}
P.~Sarmento, ``Perspectives on the {{Future}} for {{Sonic Writers}},'' \emph{Journal of Science and Technology of the Arts}, vol.~13, no.~1, pp. 110--114, 2021.

\bibitem{Meade2019}
N.~Meade, N.~Barreyre, S.~C. Lowe, and S.~Oore, ``{Exploring Conditioning for Generative Music Systems with Human-Interpretable Controls},'' Tech. Rep., 2019.

\bibitem{Sturm2016}
B.~L. Sturm, J.~F. Santos, O.~Ben-Tal, and I.~Korshunova, ``{Music Transcription Modelling and Composition Using Deep Learning},'' in \emph{Proc. on the 1st Conf. on Computer Simulation of Musical Creativity}, 2016.

\bibitem{Tan2020}
H.~H. Tan and D.~Herremans, ``{Music FaderNets: Controllable Music Generation Based On High-Level Features via Low-Level Feature Modelling},'' in \emph{Proc. of the 21st Int. Soc. for Music Information Retrieval Conf.}, Montréal, Canada, 2020, pp. 109--116.

\bibitem{Dong2018}
H.-W. Dong and Y.-H. Yang, ``{Convolutional Generative Adversarial Networks with Binary Neurons for Polyphonic Music Generation},'' in \emph{Proc. of the 19th Int. Soc. for Music Information Retrieval Conf.}, Paris, France, 2018, pp. 190--198.

\bibitem{AnnaHuang2019}
C.-Z.~A. Huang, A.~Vaswani, J.~Uszkoreit, N.~Shazeer, I.~Simon, C.~Hawthorne, A.~M. Dai, M.~D. Hoffman, M.~Dinculescu, and D.~Eck, ``{Music Transformer: Generating Music with Long-term Structure},'' in \emph{Proc. of the 7th Int. Conf. on Learning Representations}, New Orleans, LA, USA, 2019.

\bibitem{Vaswani2017}
A.~Vaswani, N.~Shazeer, N.~Parmar, J.~Uszkoreit, L.~Jones, A.~N. Gomez, {\L}.~Kaiser, and I.~Polosukhin, ``{Attention Is All You Need},'' in \emph{Proc. of the 31st Conf. on Neural Information Processing Systems}, Long Beach, CA, USA, 2017.

\bibitem{christine_2019}
\BIBentryALTinterwordspacing
C.~Payne, ``Musenet,'' 2019, {L}ast accessed: 12 Jun 2022. [Online]. Available: \url{https://openai.com/blog/musenet}
\BIBentrySTDinterwordspacing

\bibitem{Huang2020}
Y.-S. Huang and Y.-H. Yang, ``{Pop Music Transformer: Beat-based Modeling and Generation of Expressive Pop Piano Compositions},'' in \emph{Proc. of the 28th ACM Int. Conf. on Multimedia}, Seattle, WA, USA, 2020, pp. 1180--1188.

\bibitem{Sarmento2021}
P.~Sarmento, A.~Kumar, C.~Carr, Z.~Zukowski, M.~Barthet, and Y.-H. Yang, ``{DadaGP: a Dataset of Tokenized GuitarPro Songs for Sequence Models},'' in \emph{Proc. of the 22nd Int. Soc. for Music Information Retrieval Conf.}, 2021, pp. 610--618.

\bibitem{Sarmento2023}
P.~Sarmento, A.~Kumar, Y.-H. Chen, C.~Carr, Z.~Zukowski, and M.~Barthet, ``{GTR-CTRL}: {Instrument and Genre Conditioning for Guitar-Focused Music Generation with Transformers},'' in \emph{Proceedings of the EvoMUSART Conference}, 2023.

\bibitem{Adkins2023}
S.~Adkins, P.~Sarmento, and M.~Barthet, ``Looper{GP}: {A Loopable Sequence Model for {L}ive {C}oding {P}erformance using {G}uitarPro {T}ablature},'' in \emph{Proceedings of the EvoMUSART Conference}, 2023.

\bibitem{Sarmento2023ShredGP}
P.~Sarmento, A.~Kumar, D.~Xie, C.~Carr, Z.~Zukowski, and M.~Barthet, ``{ShredGP: Guitarist Style-Conditioned Tablature Generation},'' in \emph{The 16th International Symposium on Computer Music Multidisciplinary Research}, Tokyo, Japan, 2023.

\bibitem{Yang:EvaluationofMusicGen}
L.-C. Yang and A.~Lerch, ``{On the Evaluation of Generative Models in Music},'' \emph{Neural Computing and Applications}, vol.~32, 05 2020.

\bibitem{deguernel2022investigating}
K.~D{\'e}guernel, H.~Maruri-Aguilar, and B.~L.~T. Sturm, ``{Investigating the Relationship Between Liking and Belief in AI Authorship in the Context of Irish Traditional Music},'' in \emph{CREAI 2022 Workshop on Artificial Intelligence and Creativity}, 2022.

\bibitem{renPopMAGPopMusic2020}
\BIBentryALTinterwordspacing
Y.~Ren, J.~He, X.~Tan, T.~Qin, Z.~Zhao, and T.-Y. Liu, ``{{PopMAG}}: {{Pop Music Accompaniment Generation}},'' in \emph{Proceedings of the 28th {{ACM International Conference}} on {{Multimedia}}}.\hskip 1em plus 0.5em minus 0.4em\relax {ACM}, pp. 1198--1206. [Online]. Available: \url{https://dl.acm.org/doi/10.1145/3394171.3413721}
\BIBentrySTDinterwordspacing

\bibitem{daiControllableDeepMelody2021a}
S.~Dai, Z.~Jin, C.~Gomes, and R.~B. Dannenberg, ``Controllable {{Deep Melody Generation Via Hierarchical Music Structure Representation}},'' in \emph{Proc. of the 22nd {{Int}}. {{Society}} for {{Music Information Retrieval Conf}}.}

\bibitem{Dong2018a}
H.-W. Dong, W.-Y. Hsiao, L.-C. Yang, and Y.-H. Yang, ``{MuseGAN: Multi-Track Sequential Generative Adversarial Networks for Symbolic Music Generation and Accompaniment},'' in \emph{Proc. of the 32nd AAAI Conf. on Artificial Intelligence (AAAI)}, 2018.

\bibitem{luMeloFormGeneratingMelody2022}
P.~Lu, X.~Tan, B.~Yu, T.~Qin, S.~Zhao, and T.-Y. Liu, ``{{MeloForm}}: {{Generating Melody}} with {{Musical Form}} based on {{Expert Systems}} and {{Neural Networks}},'' in \emph{Proc. of the 23rd {{Int}}. {{Society}} for {{Music Information Retrieval Conf}}.}

\bibitem{wuJukeDrummerConditionalBeataware2022}
\BIBentryALTinterwordspacing
Y.-K. Wu, C.-Y. Chiu, and Y.-H. Yang, ``{{JukeDrummer}}: {{Conditional Beat-aware Audio-domain Drum Accompaniment Generation}} via {{Transformer VQ-VAE}},'' in \emph{Proc. of the 23rd {{Int}}. {{Society}} for {{Music Information Retrieval Conf}}.}\hskip 1em plus 0.5em minus 0.4em\relax {arXiv}. [Online]. Available: \url{http://arxiv.org/abs/2210.06007}
\BIBentrySTDinterwordspacing

\bibitem{jiSurveyDeepLearning2023}
\BIBentryALTinterwordspacing
S.~Ji, X.~Yang, and J.~Luo, ``A {{Survey}} on {{Deep Learning}} for {{Symbolic Music Generation}}: {{Representations}}, {{Algorithms}}, {{Evaluations}}, and {{Challenges}},'' \emph{ACM Computing Surveys}, vol.~56, no.~1, pp. 7:1--7:39. [Online]. Available: \url{https://dl.acm.org/doi/10.1145/3597493}
\BIBentrySTDinterwordspacing

\bibitem{yang2017midinet}
L.-C. Yang, S.-Y. Chou, and Y.-H. Yang, ``{MIDINet: A Convolutional Generative Adversarial Network for Symbolic-domain Music Generation},'' in \emph{International Society for Music Information Retrieval Conference}, 2017, pp. 324--331.

\bibitem{hadjeres2017deepbach}
G.~Hadjeres, F.~Pachet, and F.~Nielsen, ``{DeepBach: A Steerable Model for Bach Chorales Generation},'' in \emph{International Conference on Machine Learning}, 2017, pp. 1362--1371.

\bibitem{Yang2023Emotion}
S.~Yang, C.~N. Reed, E.~Chew, and M.~Barthet, ``{Examining Emotion Perception Agreement in Live Music Performance},'' \emph{IEEE Transactions on Affective Computing}, vol.~14, no.~2, pp. 1442--1460, 2023.

\bibitem{steinmetz2021pyloudnorm}
C.~J. Steinmetz and J.~D. Reiss, ``{pyloudnorm: {A} Simple yet Flexible Loudness Meter in Python},'' in \emph{150th AES Convention}, 2021.

\bibitem{goldsmith2014}
D.~M{\"u}llensiefen, B.~Gingras, J.~Musil, and L.~Stewart, ``{The Musicality of Non-musicians: An Index for Assessing Musical Sophistication in the General Population},'' \emph{PloS one}, vol.~9, no.~2, p. e89642, 2014.

\bibitem{braun2006using}
V.~Braun and V.~Clarke, ``{Using Thematic Analysis in Psychology},'' \emph{Qualitative research in psychology}, vol.~3, no.~2, pp. 77--101, 2006.

\bibitem{Born2020}
G.~Born, ``{Diversifying MIR: Knowledge and Real-World Challenges, and New Interdisciplinary Futures},'' \emph{Transactions of the International Society for Music Information Retrieval}, vol.~3, no.~1, pp. 193--204, 2020.

\bibitem{Gomez2013ComputationalEP}
E.~G{\'o}mez, P.~Herrera, and F.~G{\'o}mez-Mart{\'i}n, ``{Computational Ethnomusicology: Perspectives and Challenges},'' \emph{Journal of New Music Research}, vol.~42, no.~2, pp. 111--112, 2013, special issue on ‘Computational Ethnomusicology’.

\bibitem{Seeger2003IFoundIt}
A.~Seeger, ``{I found it, how can I use it? Dealing with the Ethical and Legal Constraints of Information Access},'' in \emph{International Society for Music Information Retrieval Conference}, 2003, keynote presentation.

\bibitem{sturmCopyright2019}
\BIBentryALTinterwordspacing
B.~L.~T. Sturm, M.~Iglesias, O.~Ben-Tal, M.~Miron, and E.~Gómez, ``{Artificial Intelligence and Music: Open Questions of Copyright Law and Engineering Praxis},'' \emph{Arts}, vol.~8, no.~3, 2019. [Online]. Available: \url{https://www.mdpi.com/2076-0752/8/3/115}
\BIBentrySTDinterwordspacing

\bibitem{Dong2020}
H.~W. Dong, K.~Chen, J.~McAuley, and T.~Berg-Kirkpatrick, ``{MusPY: A Toolkit for Symbolic Music Generation},'' in \emph{Proc. of the 21st Int. Soc. for Music Information Retrieval}, 2020, pp. 101--108.

\end{thebibliography}
